# Production planning in 3D Printing factories

De Antón, J.[*], Senovilla, J., González, J.M., Acebes, F., Pajares, J.

INSISOC Research Group, University of Valladolid, School of Engineering Industries, Paseo del Cauce 59, 47011 Valladolid, Spain.

*juan.anton@uva.es*

**Abstract:** Production planning in 3D printing factories brings new challenges among which the scheduling of parts to be produced stands out. A main issue is to increase the efficiency of the plant and 3D printers productivity. Planning, scheduling, and nesting in 3D printing are recurrent problems in the search for new techniques to promote the development of this technology. In this work, we address the problem for the suppliers that have to schedule their daily production. This problem is part of the LONJA3D model, a managed 3D printing market where the parts ordered by the customers are reorganized into new batches so that suppliers can optimize their production capacity. In this paper, we propose a method derived from the design of combinatorial auctions to solve the nesting problem in 3D printing. First, we propose the use of a heuristic to create potential manufacturing batches. Then, we compute the expected return for each batch. The selected batch should generate the highest income. Several experiments have been tested to validate the process. This method is a first approach to the planning problem in 3D printing and further research is proposed to improve the procedure.

## 1. Introduction

Additive Manufacturing (AM), also known as 3D printing, is a manufacturing technique that allows to produce a diversity of parts from a 3D model. The process involves the successive addition of material layers until the part is completed. In the past few years, 3D printing has evolved considerably in both techniques and materials, compared to the production processes and logistics i.e., it is evolving and is currently undergoing its phase of industrialization.

These technologies are called to play a central role in the next generation of production systems. The AM process shows benefits unachievable by traditional manufacturing techniques. Mass-customized production, prototyping, sustainable production, and minimized lead time and cost are some of the benefits that are attracting attention in the field of manufacturing (Mehrpouya et al., 2019).

As materials and techniques are constantly evolving, there is a need to develop a helpful framework to take advantage of the possibilities of AM. In production terms, this means that printing time and cost must be reduced (Gogate & Pande, 2008). Here is where the scheduling and packing problem takes on importance. Given that 3D printing a part consists of a single operation, and that multiple parts can be printed at the same time, planning and scheduling for AM brings a unique set of emerging opportunities and challenges while attempting to optimize the process (Dvorak et al., 2018). In this context, there is a growing interest to facilitate the production scheduling of AM systems.

While tackling the optimization of the 3D printing process, we can either consider the manufacturing process or the whole production cycle. The first one refers to the process of setting up the machine for the production of one or more parts, the printing of those parts and the post-processing operations. On the other hand, production planning includes all activities aimed at satisfying the order of a customer, i.e. receiving the part specifications, scheduling the part production and delivering the final product to the client.

There are different approaches to what we have referred to as production planning. In consequence, studies carried out have focused on different parts of the process. In this work we review the scheduling problem for a factory that has to schedule the daily production of several 3D printers, trying to maximize the profit obtained for each one.

Nowadays, factories rely on their know-how and intuition to set the daily production planning. In the best case, they resort to prioritization techniques backed by qualitative parameters. Both methods prove inefficient in time-saving and resource optimization. This situation demonstrates the need for an automated system that supports planning tasks. We propose the use of a heuristic to solve the planning of parts to be produced by a single machine in a daily shift.

The paper aims to provide a practical solution to 3D factories to increase their productivity by tackling the production planning. The deliverable is a Python program in which the input data of the parts is introduced and a platform layout is returned. To check their performance, the proposed heuristic will be assessed and compared to other existing techniques.

The main motivation of this paper is to provide 3D companies with mechanisms that help them increase their productivity. Having tools to optimize production capacity per unit of time leads to an increase in expected revenues per unit of time.

The article is organized as follows: Section 2 reviews the literature about planning, scheduling and nesting problems in AM. Section 3 describes the problem addressed and the main objectives. Section 4 describes the method used to solve the Packing Problem and the Winner Determination Problem, which is programmed using the software package Python. In section 5, some practical examples are presented and the outcomes are analyzed. Section 6 summarizes the main conclusions and future research is proposed.

## 2. Literature Review

Production planning in AM starts when a customer sends the design specifications of a part that needs to be manufactured (Dvorak et al., 2018). At this point, the manufacturer has to manage the printing of the part and the delivery to the customer, usually subjected to a due date. The key activity of this process turns out to be scheduling the parts that need to be printed. The scheduling process involves two fundamental questions: how to group parts for each print and how to place each part in the printing space of each print (i.e. nesting the parts) (Wang et al., 2019).

Specialists in the 3D printing field have addressed several issues related to the planning, nesting and scheduling problem. Some have acquired a comprehensive point of view of the production process, including shipping and delivery in their models. Others have focused on the scheduling and nesting problem, studying the benefits of maximizing the use of the printing surface and the simultaneous manufacture of parts. In this section, we review the most relevant works that have faced the problems previously described.

### 2.1. Production, Scheduling and Nesting Problems

The planning problem for a case in which orders from different distributed customers were satisfied by due dates was studied by Chergui et al. (2018). A mathematical formulation of the problem was presented and a heuristic was also proposed to solve it. The heuristic solution proposed was programmed in Python and tests carried out highlighted the importance of planning/scheduling for an optimized production with AM.

The nesting and scheduling problem in AM was addressed by Dvorak et al. (2018). They investigated a problem in which a set of parts with unique configurations and deadlines must be printed by a set of machines while minimizing time and satisfying both deadlines and constraints. The method proposed consisted of two main steps: first, they modeled the optimization problem of nesting parts into builds, then they scheduled those builds into machines. A similar approach was made by Wang et al. (2019) who also resorted to an analogous two-step procedure to solve the problem and presented an improved vision-based placement method.

The problem of maximizing the use of the bed (manufacturing surface) appears as one of the main issues in AM production. Kucukkoc et al.

(2016) presented a mathematical model to optimize the use of resources. However, the model was not programmed or tested to determine its performance.

The problem of production planning in AM machines was described by Li et al. (2017). A MILP model (Mixed-Integer Linear Programming) was considered and it was solved using CPLEX. In addition, two heuristics ('best fit' and 'adapted best fit') were developed for minimizing the average cost of production per volume of material. As a result, it can be affirmed that AM planning reduces costs considerably and the algorithms provided promising performance values within reasonable computing times.

According to Zhou et al. (2018) 3D printing has become a relevant service of Cloud Manufacturing, due to the vast personalization it offers to users. Cloud manufacturing provides a suitable environment for integrating the resources of 3D printing technologies. It is a platform that allows the customers to carry out their corresponding orders, satisfying the individualized requirements of material and dimensions, among others (Zhou et al., 2016). The platform groups the pieces and checks that the suppliers can do these orders.

LONJA3D project introduced by López-Paredes et al. (2018) develops a model similar to that one presented by Zhou et al. (2018), which is a Cloud Manufacturing platform. LONJA3D model is based on a platform that allows grouping the orders from various customers that meet the same material and requirements in a single order the suppliers will receive. Each supplier will have to decide which parts to do the printing. By manufacturing pieces from different clients simultaneously, the supplier manages to reduce its manufacturing costs, as shown by Piili et al. (2015), thus offering more competitive prices. Besides, the customer is favored by receiving the product at a lower price. Li et al. (2017) and Kucukkoc, (2019) tackled the chance of integrating a comprehensive nesting procedure valid to be integrated into the LONJA3D model. The allocation of parts was solved considering their exact dimensions on the horizontal and vertical axes, rather than accounting the production area as a whole. Our proposed heuristic figures out this problem, by assigning the ordered pieces to the available manufacturing area, keeping in mind to get the highest possible productivity.

Some authors have studied the benefits of packing parts in AM. Although we must note the production differences among the various AM technologies, several studies report an increase in production efficiency. Zhao et al. (2018) presented a case study where the packing algorithm showed could save over 50% of the manufacturing time compared to unpacked situations. Also, Piili et al. (2015) found that manufacturing parts simultaneously allows to reduce costs by 81-92% compared to the manufacture of parts separately, which proves that the optimal use of the manufacturing platform is the main variable that involves the supplier in terms of multi-item production.

As a case of study to demonstrate the convenience of packing parts in the printing process, Li et al. (2017) compared their novel heuristics for part scheduling ('best fit' and 'adapted best fit') with a regular assignment procedure (without a systematic approach). Results obtained with both heuristics clearly outperformed the regular procedure. As a consequence, it is important to work out an appropriate distribution of parts in manufacturing batches for reducing costs and obtaining a higher economic benefit.

## 2.2. Nesting Problem Approach

While attempting to optimize the production planning of parts from different clients, Li et al. (2017) focused on the process of assigning parts to jobs and jobs to machines (scheduling). Nevertheless, they noticed a lack of efficiency in the nesting of the pieces in the manufacturing surface. One proposed further research line was to develop a nesting method for placing parts in the manufacturing bed that could be integrated into the scheduling process. In this work, we propose a model expected to fulfill those requirements.

A part to be printed is characterized by raw material, width, length, height, volume, and filling percentage (i.e. a parameter used in AM to determine the solidness of a piece). A 3D printing manufacturer usually will have different machines to produce with different raw materials (technologies). His goal is to maximize the return with each machine, which is possible when machines are working using their whole manufacturing volume without interruption 24 hours a day, 7 days a week.

The orientation of parts is one of the most critical decisions to be made when it comes to AM. Cost, quality and time will be affected by the part's orientation. Notwithstanding, it has much to do with the technology employed; while in FDM orientation is not a key factor, in laser-based technologies it must be approached with caution (Singhal et al., 2005). Several studies have been made considering the orientation of a single part. Zhang et al. (2016) reviewed the single-part orientation problem and proposed a new model aimed at achieving the orientation of a part that allows to reduce manufacturing time and cost. Besides, an online tool was developed for users to find the right orientation according to their quality requirements.

This importance of a part's orientation lays on the anisotropic properties inherent to the layer-by-layer manufacturing process (Shaffer et al., 2014). Nevertheless, rotation around the vertical axis (Z-axis) will be allowed since the critical direction is the Z-axis itself, while orientation in the XY plane does not show a relevant influence on the final quality (Sung-Hoon et al., 2002). Vertical rotation in the nesting process has also been considered in other works, such as in Wang et al. (2019) and in Wodziak et al. (1994). This assumption allows us to increase the search space so that the possibilities of finding a better solution are higher.

The optimization problem for placing parts in the manufacturing surface has been introduced in several works with different approaches. As we have noticed for other characteristics, it has a great dependence on the 3D printing technology employed. Wodziak., (1994) prioritized the number of parts produced rather than the percentage of occupation utilized in the Stereolithography (SLA) process. Also for SLA, Canellidis et al. (2013) used three different criteria for placing the parts on the surface while the optimization objective was to maximize the area of the platform covered by the projections of the parts. Moving on to powder-bed laser processes (e.g. Selective Laser Sintering, Selective Laser Melting, Direct Metal Laser Sintering), Chergui et al. (2018) presented a model that considered delivery times in which the utilization of the manufacturing area was subordinated to the on-time service of the parts produced.

In AM the production cost (and in consequence the expected income) of a good directly depends on its mass (volume and filling percentage). To maximize the productivity of each 3D printer we should solve the puzzle that ensures for each manufacturing batch that the largest proportion of the manufacturing area will be occupied with the parts which have the highest filling percentage.

In case the amount and number of different items to be produced is large enough, with differences in mass and geometries, there is a combinatorial explosion of possible manufacturing batches. To optimize the production planning, we will consider the profit increases with the total mass of each batch.

## 3. Problem Description

3D printing factories need to be prepared to produce lots of various parts from many customers. Sometimes, the clients will order parts produced using different materials and manufacturing techniques. It opens the chance to break the lots from the clients and reorganize and combine them in case the items can be manufactured simultaneously.

We have introduced that scheduling in AM is divided into two steps: grouping parts for a print and placing parts in the printing space. In this work, we will delve into the second step, which is the most difficult to handle in computing terms. Thus, we focus on nesting parts in the manufacturing surface as a part of the scheduling process in 3D printing. The nesting problem is also referred to as platform layout optimization.

We propose a scenario in which a group of clients has made different orders of parts and the chosen supplier has to manufacture those parts. This is part of the supplier's decision making on those parts to print within the LONJA3D model. Once a large set of parts has been grouped and assigned to a printer following the matching between parts' requirements and printers' parameters, the supplier must make the decision on which parts to print in each batch. Since the capability of a printer is subjected to its printing surface, we assume that not all parts can be made at the same time. This means the supplier has to choose for the printer the parts to produce first and those that will be left for later batches.

The problem we address is the placement of parts in the bed surface once the previous grouping step has been made. A heuristic procedure is developed, in which the primary optimization criterion is to maximize the income obtained for a batch and the manufacturing surface occupation is used to help derive the final solution.

Starting from a large set of parts to be allocated in a single printer, the problem is to solve the nesting of the parts on the platform. The input data will be the width, length, height, and filling percentage of each part from the set. Also, the width, length, and height of the 3D printer will be introduced as data. The program will use those inputs to figure out an optimized layout with a subset of parts from the initial set. To solve this problem, the heuristic first seeks subsets with a large occupation percentage, and then select from those the subset showing the highest quantity of mass. Consequently, that subset and its corresponding layout would be chosen by a manufacturer that tries to schedule the daily production of that printer in an optimized way. The remaining parts will be checked on the next production planning.

### 3.1. Assumptions

We have made some assumptions in order to limit the problem to our research objective. We are relaxing several requirements to simplify the situation so that we can offer a viable solution in the shortest possible time.

A) To allocate each part in the manufacturing surface we will consider the minimum rectangle which guarantees it is inside. It is the projection of the geometry of the part on the XY plane plus the minimum tolerance needed to guarantee the quality is not compromised.

B) The location of parts is worked out in 2D, in the base formed by the X and Y axes.

C) It is assumed that all parts on the list can be manufactured by the supplier (size, precision, and cost requirements are met). The grouping step of parts whose requirements match the features of their assigned printers has already been solved.

D) Parts can rotate 90 degrees around the vertical axis.

E) Dates of delivery are not considered as well as set-up times, production times, or post-processing.

We do not explore the "matching problem" among parts and printers. It is because the goal of this work is to develop a placing procedure for a set of parts in a manufacturing surface and check the workability of our heuristic . In any case, in section 2 we have pointed out some interesting works in which matching mechanisms were presented. Thus, we will start from a list of parts to be printed and their assigned printer. In brief, we solve the problem of nesting parts from a large set in a single printer.

## 4. Batch Manufacturing Optimization Method

To maximize the productivity of 3D printing machines, we propose a method inspired by combinatorial auctions (CA). As in a CA mechanism, we will define a method with two steps. The first one is to create manufacturing batches that occupy the maximum percentage of the manufacturing bed (the Packing Problem in CA). In the second one, the Winner Determination Problem (WDP) in CA, we will evaluate for each batch the total mass (expected profit). While the most complex problem in combinatorial auctions is usually the WDP, in our case it is in the Packing Problem (PP) where the complexity is maximum.

Parts have previously been assigned to the printer with which their matching is higher. Thus, the problem to solve is the "batch composition" to determine the layout in which parts must be done to obtain an optimized solution. This problem has been solved in other works as a two-step process, in which a grouping of parts was made prior to determine their placement on the surface. Wang et al. (2019) starts from the whole list of parts to print and create a first division trying to put together parts with similar heights; then, parts of each group are divided into jobs for their successive printing. In our case, there is no previous grouping of parts and the batch composition is completely solved in the PP stage. This means that by obtaining an optimized platform layout we are solving two questions: which subset of parts must be selected and how those parts must be placed in the platform for the printing.

Optimization must be expressed in some kind of measure. This issue is here solved in the second step of the process. The winning choice will be selected according to our optimization criteria. This is what justifies the existence of a second stage in the procedure. Should we want to select another parameter as the decision variable, we would only have to modify the winning criterion.

The method will be implemented using Python. The first stage tackles the PP and solves the allocation of parts on the bed. The second deals with the winner determination problem.

### 4.1. Packing Problem

In the first stage, the PP is raised. It is started from a list of parts that will be reordered randomly in each of the simulations conducted. As we have assumed that parts will be simplified by their horizontal projections, the problem consists of finding the packing pattern that results in the largest occupied area. This problem is NP-hard because as the number of parts increases, the number of possible combinations increases exponentially. Being $n$ the number of parts, the size of the solutions space is given by:

$$2^n \cdot n! \quad (1)$$

This is a consequence of the fact that there are n! sequences of rectangles (i.e. permutations) and each one can be placed in two ways since a 90° vertical rotation is allowed. Thus, the search space is larger than the search space in the traveling salesman problem (Equation 2). If 25 parts are given, then $10^{31}$ orthogonal packing patterns exist (Jakobs, 1996).

$$2^{25} \cdot 25! > 10^{31} \quad (2)$$

We describe now the problem and the variables used for its characterization. The inputs of the parts are the name ($P_i$), filling percentage ($r_i$), length ($l_i$), width ($w_i$), height ($h_i$). The build platform area is given by name ($A_j$), length ($L_j$), width ($W_j$) and height ($H_j$), all of them in mm. We introduce the variable $x_{ij}$ as the Boolean variable that takes the value 1 if the part is assigned to the batch. Also, we define two variables that indicate the position of the part on the XY plane by the top-left corner on the X-axis ($cX$) and the top-left corner on the Y-axis ($cY$).

First, we introduce two new terms that will be used through the algorithm: the List of Available Parts (LAP), which contains the parts that remain to be assigned, and the List of Available Areas (LAA), where a list of the unused areas can be found.

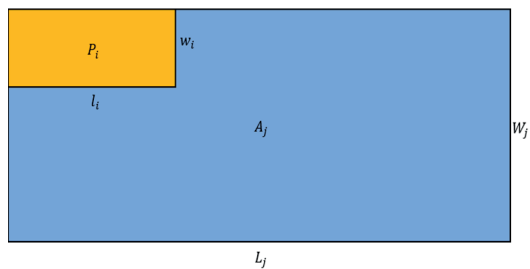

**Figure 1.** Placement of the part $P_i$ on the area $A_j$.

The procedure begins with an empty surface. The first part will be allocated in the upper-left corner. Then, two new subareas will be created from the remaining available surface: the first one by cutting from top to bottom, and the second one covering the remaining area.

To start the allocation procedure, we choose the first part of the LAP and try to assign it to the first area of the LAA. To do this, we follow several steps (see Figure 2):

1. We compare the width of the part ($w_i$) and the width of the area ($W_j$).

   a. $w_i < W_j$, then lengths are compared

      i. $l_i < L_j$, then the part is assigned and $x_{ij}$ takes the value 1. Also, we create two new areas from the remaining available surface (see Figures 3 and 4). The first one, ($A_{j+1}$), is defined by:

      $$A_{j+1} = (l_i, W_j - w_i) \quad (3)$$

      The length of the new area coincides with the length of the part assigned. The width is calculated as the difference between the width of the original area ($A_j$) and the width of the part. Similarly, we define ($A_{j+2}$) by:

      $$A_{j+2} = (L_j - l_i, W_j) \quad (4)$$

      These two new areas are included in the LAA, while the area $A_j$ is removed from the list. To define the position of these areas, they will be assigned their corresponding pair of coordinates from those variables defined at the beginning ($cX$ and $cY$).

      ii. $l_i > L_j$, we rotate the part 90° so that the new width $w_i'$ coincides with the former length and the new length $l_i'$ coincides with the former width. Once the change has been made, we repeat the checking procedure:

   b. $w_i' < W_j$, then the length is checked

      ii. $l_i' < L_j$, then the part is assigned. Two new subareas are created as explained in (i) and the LAA is updated.

   c. $w_i' > W_j$, we move on to the next area of the LAA and restart from the first step.

2. $w_i > W_j$, we rotate the part 90° as we have explained in (ii) and repeat the checking procedure.

A second piece follows the same procedure, i.e. if it does not fit in the first subarea, the second created subarea is checked for compatibility. This procedure is iterated until there are not available parts or surfaces that can be assigned.

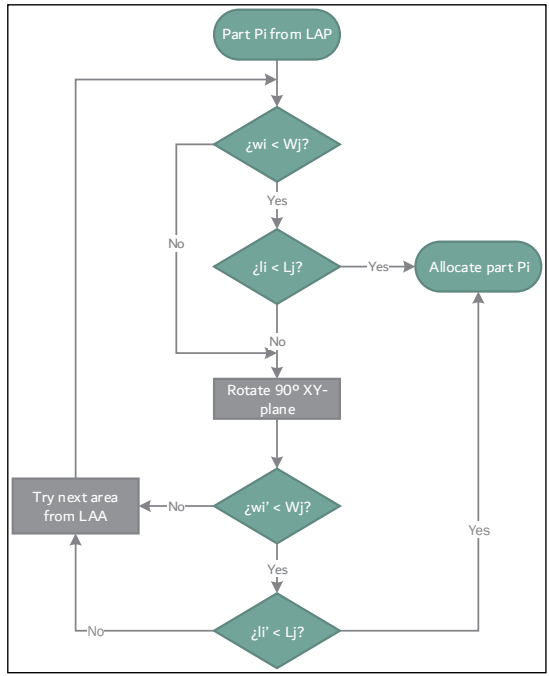

**Figure 2.** Flow chart for the allocation of a part.

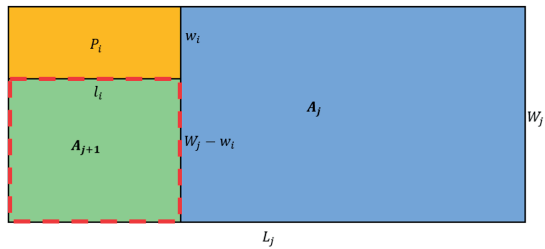

**Figure 3.** Dimensions of the new area $A_{j+1}$.

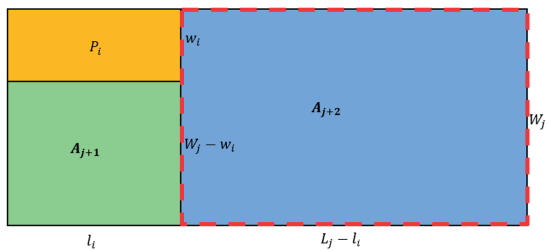

**Figure 4.** Dimensions of the new area $A_{j+2}$.

### 4.2. Winner Determination Problem (WDP)

As well as other authors such as Li et al. (2017) try to minimize the average production cost per volume of material, and others try to minimize the tardiness in the delivery of produced parts, see Dvorak et al. (2018) and Chergui et al. (2018), we acquire an alternative perspective of the planning problem: to maximize the income we get from each printer. This focus has its support on the fact that the goal of a company is to increase the profit, and resources and management systems are all subordinated to that goal (Gupta & Boyd, 2008). Also, this approach runs in concordance with our scenario: do the daily planning for a 3D factory.

Once the manufacturing surface has been optimized (PP), the supplier has to choose among all the possible batches the one that offers the highest income. Despite having solved the PP in 2D, parts are also characterized by their height ($h_i$) and filling percentage ($r_i$). The profit generated by each part has a close relationship with those two parameters and a decision criterion must be set to select the winner batch among those showing the highest occupancy percentage.

One time the PP step finishes, the program provides two outputs that help determine the winner choice. The first one provides the given occupied manufacturing area. The second gives the total mass used for each batch. The last one allows us to solve the winner determination problem.

Parts have been simplified by their horizontal projections in 2D, and to determine their volume we will consider them as cubes. Then, the volume of a part ($v_i$) is given by:

$$v_i = l_i \cdot w_i \cdot h_i \qquad (5)$$

In order to calculate the amount of material needed to manufacture a part ($m_i$), we simply multiply the volume of the cube by the filling percentage ($r_i$) as follows:

$$m_i = v_i \cdot r_i \qquad (6)$$

The total stuff of the batch is obtained by adding up the single masses of the parts assigned to it, and it is expressed as a volume (in mm³). This will allow to choose the batch that has the largest amount of stuff intending to obtain the highest income, being this a useful tool for the provider to maximize the return of the machine.

Costs in AM can be simplified as a function of two components: a mass-dependent one and a constant that represents the pre-processing and post-processing costs.

$$C(m) = k + c \cdot m \qquad (7)$$

At the same time, material costs constitute a major proportion of the costs involved in producing a part with AM techniques (Thomas & Gilbert, 2015). Though this makes more sense for big parts rather than small ones (where the material cost is less relevant as compared to other AM costs), we will assume that in our model it occurs indifferently. On an analogous reasoning, the income derived from the production of a part with AM is mainly represented by the mass of the part multiplied by the price per kg. Being I the income expected:

$$I(m) = p \cdot m \qquad (8)$$

This is how we justify our winning choice as the batch with the highest mass.

## 5. Experimentation: Results

A case of study has been used to validate the method. It is considered a 3D printer whose parameters are shown in Table 1.

In addition, 10 parts have been listed (see Table 2). These meet the requirements already described. 120 simulations are done to derive a solution close to the optimal solution. The outputs of the simulations are displayed in Tables 3 and 4. The results showed that in the simulation 77 the batch composed by parts P1, P9, P4, P7, P5, P2, and P8 got an occupation of the manufacturing area of 92.69% (37 075 mm$^2$ of the 40 000 mm$^2$ available). The mass used was 1 523 500.00 mm$^3$.

In simulation 26, it is obtained a batch with a higher percentage of occupation. The batch consists of the pieces P2, P7, P1, P3, P5, P6, P8 (see Figure 5) with an occupancy rate of 97.63% (39 050 mm$^2$ out of the 40 000 mm$^2$ available). It is approximately 5% higher than the previous case. However, the amount of raw material is 1 502 500.00 mm$^3$. That is a 21 000 mm$^3$ difference from the previous case. As a consequence, the first option is selected as the winner since it should generate the greatest benefit for the manufacturer.

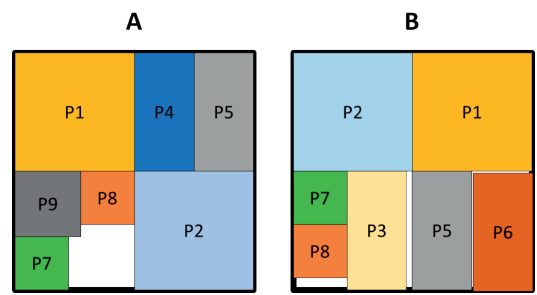

**Figure 5.** A) A batch that occupies the 92.69%. B) A batch which occupies the 97.63%.

These results show that our heuristic achieves good performance in absolute terms. However, we will carry out some experiments to have a more accurate idea of how precise our approach is. We have addressed the Packing Problem and the Winner Determination Problem separately, and now we introduce the findings reached.

As for the PP, the two main concerns are the order of the parts in the list (LAP) and the performance of our procedure compared to other authors' studies. It seems reasonable to think that the order of parts to enter the algorithm could have an effect on the solution obtained.

Thus, we conduct an experiment forcing the order of the parts by size: first, we try allocating parts ordered from largest to smallest area; second, we try with parts ordered from smallest to largest area. Results obtained reported that our heuristic performs better in the first case but compared to a random order of the parts it still shows a lower performance (see Table 5). In conclusion, we can say that integrating local optimal search techniques can help save computing time and guide the solution.

To check the performance of our heuristic, we have conducted experiments where we compared the solutions reached with those obtained by other authors. Two scenarios are analyzed: a case with few parts and a case with a large number of parts to allocate. We run several experiments and, as an example, we will discuss the results obtained for an experiment with a list of 40 parts of 10 types and one with 106 parts of 20 types. Because we are studying the PP, we analyze the results of surface occupation.

For the case with low parts we use an experiment conducted by Toro & Granada-Echeverri, (2007) since they include the geometric data of the parts.

**Table 1.** Features of the manufacturing area.

| Name | Length(mm) | Width (mm) | Height (mm) | Area (mm$^2$) | Volume(mm$^3$) |
|---|---|---|---|---|---|
| A1 | 200 | 200 | 200 | 40 000 | 8 000 000 |

**Table 2.** Attributes of parts.

| Name | Length (mm) | Width (mm) | Height (mm) | Filling | Area (mm$^2$) | Volume(mm$^3$) |
|---|---|---|---|---|---|---|
| P1 | 100 | 100 | 100 | 0.5 | 1 000 000 | 500 000 |
| P2 | 100 | 100 | 100 | 0.5 | 1 000 000 | 500 000 |
| P3 | 50 | 100 | 100 | 0.2 | 500 000 | 100 000 |
| P4 | 50 | 100 | 100 | 0.2 | 500 000 | 100 000 |
| P5 | 50 | 100 | 100 | 0.2 | 500 000 | 100 000 |
| P6 | 50 | 100 | 100 | 0.2 | 500 000 | 100 000 |
| P7 | 45 | 45 | 100 | 0.5 | 202 500 | 101 250 |
| P8 | 45 | 45 | 100 | 0.5 | 202 500 | 101 250 |
| P9 | 55 | 55 | 100 | 0.4 | 302 500 | 121 000 |
| P10 | 80 | 80 | 100 | 0.3 | 640 000 | 192 000 |

**Table 3.** Batch associated with the largest amount of stuff.

| Name | Length | Width | Height | Filling | Area (mm$^2$) | Volume(mm$^3$) | Stuff (mm$^3$) |
|---|---|---|---|---|---|---|---|
| P1 | 100 | 100 | 100 | 0.5 | 10 000 | 1 000 000 | 500 000 |
| P9 | 55 | 55 | 100 | 0.4 | 3 025 | 302 500 | 121 000 |
| P4 | 50 | 100 | 100 | 0.2 | 5 000 | 500 000 | 100 000 |
| P7 | 45 | 45 | 100 | 0.5 | 2 025 | 202 500 | 101 250 |
| P5 | 50 | 100 | 100 | 0.2 | 5 000 | 500 000 | 100 000 |
| P2 | 100 | 100 | 100 | 0.5 | 10 000 | 1 000 000 | 500 000 |
| P8 | 45 | 45 | 100 | 0.5 | 2 025 | 202 500 | 101 250 |
| | | | | Total | 37 075 | 3 707 500 | 1 523 500 |

**Table 4.** Batch with the highest percentage of the area covered.

| Name | Length | Width | Height | Filling | Area (mm$^2$) | Volume(mm$^3$) | Stuff (mm$^3$) |
|---|---|---|---|---|---|---|---|
| P2 | 100 | 100 | 100 | 0.5 | 10 000 | 1 000 000 | 500 000 |
| P7 | 45 | 45 | 100 | 0.5 | 2 025 | 202 500 | 101 250 |
| P1 | 100 | 100 | 100 | 0.5 | 10 000 | 1 000 000 | 500 000 |
| P3 | 50 | 100 | 100 | 0.2 | 5 000 | 500 000 | 100 000 |
| P5 | 50 | 100 | 100 | 0.2 | 5 000 | 500 000 | 100 000 |
| P6 | 50 | 100 | 100 | 0.2 | 5 000 | 500 000 | 100 000 |
| P8 | 45 | 45 | 100 | 0.5 | 2 025 | 202 500 | 101 250 |
| | | | | Total | 39 050 | 3 905 000 | 1 502 500 |

**Table 5.** Comparison of the performance for the initial order of parts in the LAP.

| Order | Parts allocated | % of area covered |
|---|---|---|
| Largest - Smallest | P4, P5, P6, P7, P8, P12, P13, P14 & P15 | 88,19 |
| Smallest - Largest | P4, P5, P6, P9, P10, P11, P13 & P14 | 44,44 |
| Random | P1, P2, P3, P4, P6, P7, P8, P9, P10, P11, P13 & P14 | 100 |

To solve the problem, they use the Chu-Beasley genetic algorithm and run it for 28 seconds. The best solution found reaches an occupancy percentage of 95%. Running our heuristic for the same time we achieve 100% of occupation.

The experiment with a list of 106 parts is introduced by Cui, (2007) and we compare our results with those obtained by Toro et al. (2008). The search space for this problem is up to $2^{106} \cdot 106! = 9.2997 \cdot 10^{201}$ solutions, so here the complexity is high. They propose a complex algorithm that merges different heuristic techniques (e.g. neighborhood search, simulated annealing) to find the best solution, which shows occupancy of 99.31%. We make 100 000 simulations with our heuristic and the best solution found reaches 94.03%.

The conclusion reached for those experiments is that our heuristic performs well for a case with a relatively small number of parts, outperforming in some cases the results reported in the existing literature. Nevertheless, as the complexity of the problem increases with the increase in the number of parts, the results are not as good as others analyzed, and computing time increases considerably.

Attending to the winner determination problem, we are pursuing the batch that has the highest amount of stuff. The two decisive parameters in this context are height and filling percentage. We conduct experiments to study their significance degree. In both studies we analyze a case in which we start from a list of 15 parts to allocate; we realize 10 000 simulations and choose the 8 best solutions.

To study the importance of the filling percentage we consider that all parts have the same height. We compute the algorithm and analyze the results (see Table 6). We see that the batch with the highest amount of stuff (i.e. the winner) is not the one with a higher percentage of area covered neither the one with the highest number of parts allocated. This can be explained because here the total mass is primarily determined by the filling percentage. In conclusion, we can say that parts with a high filling percentage have more value.

**Table 6.** Experiment with parts of the same height.

| Solution | % area covered | Nº parts | Stuff (mm$^3$) |
|---|---|---|---|
| Winner | 97.22 | 11 | 20 800 000 |
| Max area | 100 | 12 | 20 300 000 |
| Max nº parts | 98.61 | 13 | 20 200 000 |

Alternatively, we study the case keeping the filling percentage constant (and varying the heights) to analyze the significance of the height. As it happened in the previous study, the batch with the highest amount of stuff has less parts allocated and less area occupied than other solutions (see Table 7). This allows us to prove that height has an important influence on the WDP and that parts with higher height can bring greater benefits to the detriment of others with greater area.

**Table 7.** Experiment with parts of the same filling percentage.

| Solution | % area covered | Nº parts | Stuff (mm$^3$) |
|---|---|---|---|
| Winner | 95.14 | 11 | 79 500 000 |
| Max area | 100 | 12 | 75 000 000 |
| Max nº parts | 98.61 | 12 | 76 750 000 |

## 6. Conclusions and Future Research

Production optimization can be understood in different ways. The approach taken in this paper is that suppliers should choose the batch that allows them to get the highest income. In a dynamic market, where customers are continuously making new orders, this allows to choose those parts that bring the maximum return at the end of the day. At the same time, the main problem is to solve the nesting of parts on the manufacturing surface trying to maximize the occupied area.

A two-step method based on combinatorial auction has been presented. First, we solve the PP to create potential manufacturing batches. Then we calculate the expected return for each one (the WDP).

After conducting several experiments in which we compared our method with other techniques previously developed, we have reached interesting conclusions about its performance. We found that our program offers a better result when parts in the list are sorted from largest to smallest area. Also, when the number of parts to allocate stays around 40, we obtain a quality solution in 5-10 minutes of computing time. However, for a list of about 100 parts the solutions show a lower quality and the computing time fires above 15 minutes. Another finding proved was that the filling percentage and the height are relevant attributes, forcing to consider situations with a lower occupation of the bed but a higher use of material.

Once the conclusions have been discussed, we focus on the research lines that will enable the improvement of our procedure. We know that our method does better when parts are sorted from large to small. Notwithstanding, parts in the list are randomly sorted. This opens up the possibility of integrating new search techniques to explore the neighborhood of the best solutions obtained in each simulation. With regards to the height, we have highlighted its importance in our WDP. Nevertheless, differences in height among parts will probably translate into a waste of time in multi-parts manufacturing, since a short part cannot be removed from the plate until the tallest part is finished. A good solution could be a penalty factor for parts with noticeable differences in height.

The time dimension is not included in our method. A main research line can be the integration of algorithms to calculate the time needed to manufacture a batch. At the same time, it will be interesting to expand this study (designed for a single machine) to a multi-machine production scheduling.

By integrating both problems (PP and WDP) in a comprehensive approach, we would need to introduce a bounding criterion to select a subset of the batches with a higher mass. This might be interesting according to our aim of choosing the batch with the highest mass. However, the searching process proves to be more flexible when setting the area covered as the first filtering parameter. Besides, in the outlook of improving this procedure, we feel it is preferable to figure out both problems separately so that they can be addressed independently to improve their performance.

## Acknowledgements

This research has been partially financed by the project: "Lonja de Impresión 3D para la Industria 4.0 y la Empresa Digital (LONJA3D)" funded by the Regional Government of Castile and Leon and the European Regional Development Fund (ERDF, FEDER) with grant VA049P17.